# Understanding Security Requirements and Challenges in Internet of Things (IoTs): A Review

**Faraz Idris Khan and Sufian Hameed**
National University of Computer and Emerging Sciences (NUCES)
Department of Computer Science
IT Security Labs
Karachi, Pakistan
[e-mail: faraz.idris@nu.edu.pk, sufian.hameed@nu.edu.pk]
*Corresponding author: Sufian Hameed

## Abstract

Internet of Things (IoT) is realized by the idea of free flow of information amongst various low power embedded devices that use Internet to communicate with one another. It is predicted that the IoT will be widely deployed and it will find applicability in various domains of life. Demands of IoT have lately attracted huge attention and organizations are excited about the business value of the data that will be generated by the IoT paradigm. On the other hand, IoT have various security and privacy concerns for the end users that limit its proliferation. In this paper we have identified, categorized and discussed various security challenges and state of the art efforts to resolve these challenges.

## 1. Introduction

The emerging trends in embedded technologies and Internet have enabled objects surrounding us to be interconnected with each other. We envision a future where IoT devices will be invisibly embedded in the environment around us and it will be generating enormous data. This data have to be saved and processed in the end to be presented in an understandable form.

IoT model involves numerous actors which includes mobile operators, software developers, access technology providers. Additionally, IoT involves numerous different application domains i.e. manufacturing, utility management, agriculture and healthcare. IoT can be seen as the next generation interconnection paradigm which will enable connectivity among people's devices and machines enabling actions to happen without human intervention. The success of the IoT world requires merger of different communication infrastructure. This has lead to the design of smart gateways to connect IoT devices with the traditional Internet. Most recently efforts are directed to interconnect IoT infrastructure and Cloud Computing which supplements the potentials of IoT.

Increasing complexity of IoT networks also magnifies the number of security challenges. This is attributed to huge amount of devices connected to the Internet along with huge data generated by these devices which is a major concern for security. Attacks in IoT is possible as the devices in IoT network become easy targets for intrusion [1]. Upon compromise the hackers can gain control and do malicious activity on the device and attack other devices close by the original compromised node. IoT devices do not have virus protection or malware protection which makes them highly susceptible to become bots and carry out malicious activity to other devices in the network. Also, by hijacking the IoT device the routing and forwarding operations can also be sabotaged by the malicious users. In addition to attacks on IoT devices, personal information gathered by the IoT devices can be compromised as a result of cyber-attack or unauthorized access. This lack of confidentiality, integrity and security of data in IoT will make IoT applications less popular [2]. Problem of securing IoT devices is further aggravated due to their resource constrained nature, due to which they have limitations of energy and computational resources.

In this paper, we have discussed the state of the art efforts in securing IoT network and applications under the security challenges briefly highlighted above. These challenges mainly fall under privacy in IoT, lightweight cryptographic framework for IoT, secure routing and forwarding in IoT, robustness and resilience management in IoT and DoS and Insider attack detection in IoT. Furthermore, we have identified and discussed open issues and challenges in each of the state of the art efforts discussed under these categories.

The rest of the paper is organized as follows. Section 2. discusses privacy issues in IoT. In section 3, state of the art in proposing lightweight cryptographic framework for IoT are discussed. Section 4 discusses all the state of art proposals in secure routing and forwarding for IoT. State of the art in provisioning resilience and robustness management in IoT are discussed in section 5. Section 6 summarizes state of the art in proposed denial of service and insider detection in IoT. We conclude our paper in section 7.

## 2. Privacy in IoT

### 2.1 Motivation

Privacy in IoT is a prime security issue that needs full attention from researchers in academia and industry. There is a desire need to propose protocols and management framework for handling privacy in IoT. IoT have become an integral part in various applications like remote patient monitoring, energy consumption control, traffic control, smart parking system etc. In all of them users require protection of personal information which are related to its movement, habits and interactions with other people.

### 2.2 Challenges

Association of an identity to a certain individual is a threat as this may lead to profiling and tracking. Hence, one of the major challenge is to disallow such activity in IoT and take some preventive measures. Localization and tracking is threat to determine and record person's location through time and space. Major challenge lies in designing protocols for interactions with IoT that discourages such activity. Profiling information related to a certain individual to infer interests by correlation with other profiles and data is very common in e-commerce applications. Huge challenge lies in balancing interests of businesses for profiling and data analysis with user's privacy requirements. Other challenges of ensuring privacy in IoT includes; transmitting data in a secure manner through a public medium without concealing to the general participants in the network, preventing unauthorized collection of information about existence and characteristics of personal things.

### 2.3 Existing Solution and Discussion

In this section, we discuss existing efforts in the direction of ensuring privacy in IoT application especially Body Sensor Networks.

Most recent work which addresses security and privacy challenges of cloud based IoT can be found in [3]. Security and privacy requirements in cloud based IoT as identified by the authors are identity privacy, location privacy, node compromise attack, layer removing/adding attack, forward and backward security, semi trusted and malicious cloud security. Another recent work that is an attempt to analyze existing privacy preserving solutions can be found in [4]. The authors identified the gaps in various proposals and put forward suggestions to remove them.

The authors in [5] surveyed existing IoT applications. In this work, authors proposed a translation of their modules in a common system model and at the same time identified and studied differentiating behavioral pattern of sensor data generated. From the analysis, it was disclosed that almost all applications gather location and time information. Whatever data is gathered can be of various types including video and audio. The authors surveyed up to date privacy countermeasures. Furthermore, potential threats to user privacy in participatory sensing which result from uncontrolled disclosure of personal information to untrusted people have been discussed. Also, the authors mapped their analysis to a proposed common system model for analyzing security in mobile participatory sensing application.

A detail discussion on security threats and privacy in IoT architectures can be found in [6]. The discussion begins with detail layered architecture of IoT. Privacy and security threats at each level of the architecture is analyzed in detail. State of the art in presenting threat scenarios at various levels of the IoT architecture is discussed in detail. Based on the scenarios discussed the security issues of importance are eavesdropping, man-in-the-middle and other similar attacks that jeopardize the data confidentiality and integrity, and grabbing control of some components. Along with that the authors also study the emerging EU legislation for IoT. It is important to understand management domains of IoT architecture. EU legislation requires an individual should be able to control his or her information at all levels of architecture. Issues of further study requires in depth study of how this kind of control is technically supported. Energy aspects of privacy and threats require more in depth study.

In [7], the authors surveyed privacy enhancements in IoT in various application domains. Key future security requirements for smart home systems are discussed. Also, the authors suggested a suitable security architecture for IoT. Gateway architecture is nominated to be the most appropriate for resource-constrained devices and for high system availability. This architecture implement sophisticated management algorithms on a reasonably powerful processor and can operate critical smart home functions. Apart from gateway architecture other architectures are scrutinized for IoT are middleware architecture and cloud architecture. Two technologies are discussed for a gateway architecture for assisting auto management. Firstly, auto configuration support enhancing system security. Secondly, automatic update of system software and firmware to maintain ongoing secure system operation.

Efforts in managing privacy for IoT by efficient data tagging through IFC (Information Flow Control) tags can be found in [8]. The sensed data is tagged with privacy properties that allows trusted control access based on sensitivity. Due to resource requirement of IoTs tagging is very expensive and this work discusses the concerns about tagging resource constrained IoT. Four properties of privacy sensitive IoT applications includes physical interaction, sensing valuable data, distributed implementation and vulnerable sensors are illuminated which makes IFC data tag feasible for privacy preservation. Apart from these four there are two more properties of such applications that are connected operation and skewed tag use that makes implementation of IFC data tags much easier.

In [9], security challenges in mobile adhoc networks are discussed in detail. There are various challenges to security design such as open peer to peer network architecture, shared wireless medium, stringent resource constraints and highly dynamic network topology. Considering these challenges building a multi-fence security solution that achieve both broad protection and desirable network performance is possible. Security issues and state of the art proposals related to multihop delivery of packets among mobile nodes is discussed. For a comprehensive security solution, it should span both layers and encompass all three security components of prevention, detection and reaction.

Enabling technologies for IoT privacy provisioning such as RFID can be found in [10] with detail discussion on threat analysis of RFID system components. RFID technology is considered good for tracking and keeping stock of items. In order to apply this to humans there have to be laws, regulation to operate, and strict imposition to ensure acceptance as it can be abused. The authors conclude in order to use RFID to enable IoT, issues with technological and social problems have to be resolved

In [11] authors have proposed a Host Identity Protocol (HIP) and Multimedia Internet Keying protocols enabling secure network alliance with the network in a secure manner along with managing keys using key management mechanism. HIP leverages public key cryptography to provide distinct identification of the IoT devices. Furthermore, the authors have extended HIP to have key management support.

Medical Sensor Networks (MSN) requires efficient and reliable access control that is a crucial requirement to authorize staff to access private medical data and ensure productive and dependable access control. The authors in [12] have proposed an access control system enabling control on access in well-defined medical situations. Proposed system is an extension of the modular traditional role based access control model. Modular design enables simpler way of making decision for access control and effective distribution of access control policies.

In [13] the authors proposed a privacy protection mechanism based on a concept of path jumbling which is a collective privacy preserving mechanism. With the proposed mechanism, user's privacy is preserved in a redistributed fashion by exchanging sensor readings.

The authors in [14] propose a detailed analysis of threats related to privacy challenges in the IoTs. Detail analysis of seven threat categories identified by the authors is discussed. These seven categories are identification, localization and tracking, profiling, privacy violation interaction and presentation, lifecycle transitions, inventory attack and linkage. Identification is a threat of attaching a (persistent) identifier with an individual and data about him. Location and tracking is threat of determining and recording a person's location through time and space. Profiling is a threat of compiling information of individuals that infer interests by correlating with other profiles and data. Privacy violating interaction is a threat of conveying private information by a public medium and disclosing it to unwanted audience. Changes of control sphere during lifecycle transition threatens privacy as smart things disclose private information. Unauthorized collection of information about the existence and characteristics of personal things is called inventory attack. In addition, another privacy threat is associated in combining and aggregating data from different data sources which often happen in IoT application. This will reveal information from a single data source that is not intended to be made public when its data was isolated. The authors conclude that profiling remains one of the severe threat that needs attention from the research community. The authors discuss two core thoughts firstly, IoT is evolving which makes privacy a constant challenge and secondly, a comprehensive framework is required, which caters threats identified by the authors.

In [15], authors have discussed security of mobile sensing application. In this work, the authors presented nominated application scenarios in order to spotlight potential benefits inferred from their utilization. Authors studied heterogeneous statistics acquired by current mobile sensing applications and the flow of data in the application architectures. Particular emphasis is given on threats related to privacy. These threats are on primary information and sensor readings collected in existing mobile sensing deployments. These readings are spatio-temporal information, sound samples, pictures, videos, and accelerometer data. Temporal annotation of sensor data provide insights about the habits of the users and endanger privacy. Spatio-temporal information threatens privacy of the users in their own. Automated recording of sound samples poses serious risks for user privacy in the absence of privacy preserving mechanisms as confidential conversation can be recorded. Pictures and videos endanger privacy of additional people captured in images by the user. As it may reveal their current location as well as identity of social relations.

The authors in [16] have presented security architecture for IP enabled IoT based on HIP (Host Identity Protocol) and DTLS ((Datagram Transport Layer Security) adapted to resource constrained devices. In addition, authors propose key management architecture for IoT. Privacy protection is provided by proposing HIP-PSK (Host Identity Protocol - pre-shared keys ) and DTLS -PSK (Datagram Transport Layer Security - pre-shared keys ) that provide secure network access and communication.

In [17] the authors identified and discussed three representative sensing applications. They are personal sensing, designated sensing and community sensing each requiring heterogeneous security and privacy guarantees. Wireless Community Networks (WCN) realizes these sensing applications. WCN is emerged from the integration of wireless mesh networks, wireless sensor networks and mobile communications that will form the future communication infrastructure for urban communities. Heterogeneous sensing applications have different security and privacy challenges. These security challenges are motivated by integration of sensors into WCN infrastructure. Such challenges are raised due to presence of sensors in houses of community members or in mobile devices. First challenge is to devise security and privacy model in order to understand exact risks and threats for each of application types. Second challenge is in personal and designated sensing applications. The access to measurements should be based on appropriate privacy preserving access control mechanism. Third challenge is to device mechanisms for privacy preserving community sensing. Where the main concern is related to the anonymized sensed environmental data which could become public. Moreover, [18] proves to be useful in implementing privacy in sensing applications as suggested by the authors in [17].

In [19] the authors categorized approaches in tackling IoT as rule based and architectural based approaches. They propose architecture based privacy protection framework. IoT is modelled as Cooperative Distributed Systems where things cooperate to attain individual or cooperative goals. Contract Net Protocol (CNP) is extended to support privacy protection for IoT.

Latest evaluations of IoT networks for security and privacy can be found in [20]. The authors focused on preserving privacy in home automation networks that is claimed to be extended to IoT applications. It is demonstrated that both basic cryptographic techniques and data manipulation are employed to save a user against a rival inside the IoT network or rival who have compromised remote servers.

A comprehensive survey of privacy and trust issues in IoT can be found in [21]. Traditional security countermeasures are quite different and cannot be applied immediately to IoT technologies due to heterogeneous standards and communication stacks involved. There is a desire for a flexible architecture to deal with hazards in dynamic environment where scalability issues arises due to high number of interconnected devices. The authors presented and discussed essential research challenges and prevailing solutions in IoT security, distinguishing open issues and proposing future directions for research. Besides, challenges in IoT is also discussed in [22] where authors introduced industrial IoT and discussed relevant security and privacy challenges and further give viable solutions which leads to comprehensive security framework. More summarization of security threats and privacy concerns of IoT can be found in [23]. There are efforts in establishing relation between information, privacy and trust that can be found in [24].

In [22] analysis of security requirements, threat models and security issues in IoT is discussed in detail with a comprehensive classification and taxonomy of attacks. Open problems and latest research issues are discussed in the paper. Possibilities for future research work is also discussed in the paper. Latest effort in classifying security in IoT can be found in [25]. The authors proposed a model based on privacy and classification called privacy information security classification (PISC). Privacy is divided into three security levels. Level 1 privacy is related with public information leakage of which will not cause serious consequences to the owner. Level 2 privacy is compose of anonymous and semi anonymous personal data. Level 3 privacy is information directly corresponding to user's identity such as fingerprint, identification card information and internet protocol address. Forging of fingerprint leads to stealth of useful information. Therefore, complex security protection measures in protecting level 3 privacy information is required. Different security protection technologies are required to achieve different security goals for different level of privacy information.

Research community have proposed protocols for ensuring privacy in IoT such as in [26] the authors proposed two sensor based secure communication protocol for healthcare systems based on IoT. Latest work on key management protocol for IoT can be found in [27]. The protocol also performs robust key negotiation, lightweight node authentication, fast rekeying and efficient protection against replay attacks. Proposed key management protocol (KMP) is integrated at layer 2 of protocol stack. It leverages "fixed" Elliptic Curve Diffie Hellman (ECDH) exchange and the (Elliptic Curve) Qu-Vanstone implicit certificates. KMP is implemented in open source OpenWSN protocol stack. A comprehensive survey on secure communication protocols for IoT can be found in [28] where the authors have highlighted the inapplicability of traditional security protocols in IoT and give a detail taxonomy of key distribution and management protocols.

## 2.4 Open Issues and Future Directions

In all the existing work that exist in literature till now there is no comprehensive framework that ensures privacy in IoT for a large class of applications. There is a desire need to have a generic lightweight cryptographic privacy-preserving algorithm that ensures confidential exchange of data at the same time anonymizing the origin of data.

Latest trends in IoT privacy protection are user-centric and context-aware privacy policies. Along with them, other emerging techniques are context-centric and self-adaptive privacy preserving mechanisms and protocols supporting ambient intelligence. One of the novel emerging field is of privacy preservation of data streams in IoT. This requires dynamic data access control mechanisms and data management policies.

Game theory has lately been used to analyse location privacy. An interesting open research question is how to implement incentives in the IoT architecture's privacy preserving protocols by utilizing game theory. For the next generation networks context management is expected to interact with underlying IoT technologies and deal with related privacy issues improving quality of the context.

Hence, for the coming years development of more sophisticated privacy models and practically relevant privacy oriented security protocols and mechanisms are identified as research direction of extreme importance. Network virtualization adaptation for preserving privacy for huge amount of data being handled in IoT deployments and cloud management has emerged as a potential approach. Software Defined Networking (SDN) lately has emerged as a paradigm for network virtualization. Also, SDN regulates the network by centralizing the routing and forwarding functionality at a central point, which is known as the controller. This can help the network operator and administrator to implement privacy over the whole network. Initial architecture in this direction be found in [29].

## 3. Lightweight Cryptographic Framework for IoT

### 3.1 Motivation

IoT introduces new challenges in terms of energy and power consumption. It is desired that the cryptographic primitives designed for IoT should be lightweight. These primitives must consume fewer resources without compromising the required level of security. Hence, research community have started focusing on lightweight cryptography. Properties of lightweight cryptography is discussed in ISO/IEC 29192, ISO/IEC JTC 1/SC 27. There is also a project of lightweight cryptography (ISO/IEC 29192) under the process of standardization. Lightweight cryptography in ISO/IEC 29192 is described based on target platform. Chip size and energy consumption are important measures to assess lightweight properties. Further, small code and/or RAM size are preferable for the lightweight applications in case of software implementation.

### 3.2 Challenges

Given the constraints of hardware resources there is a desire need to design lightweight cryptographic framework for IoT. This can be achieved by proposing cryptographic primitives that needs to be revisited and designed considering the constraints of IoT devices.

## 3.3 Existing Solution and Discussion

In this section, we discuss efforts in the direction of proposing lightweight cryptographic framework for IoT.

The authors in [30] have proposed a security architecture that confirms the security goals discussed in the paper. Proposed solution works according to the lifetime of a smart object in IoT network. The keying material are managed by TTP infrastructure. The proposed framework is used to make manufacturing of smart objects in a protected manner.

In [18] the authors have proposed a resource friendly, fast and distributed security mechanism for key agreement and verification of identification parameters in WSN. Proposed system is based on alpha secure polynomials that has been proposed for key distribution and establishment. The authors in [31] have proposed mechanisms to make the computation of polynomials more lightweight for IoT.

The authors in [32] have identified three devastating security compromise, initiated from the Internet, at transport layer in opposition to the low power and lossy networks. The authors observe provisioning E2E security is not possible with ease due to variety of usage scenarios like CoAP/CoAP ( Constrained Application Protocol ), DTLS/DTLS (Datagram Transport Layer Security), HTTP/CoAP, TLS/DTLS which are arbitrated by 6LBR (6LoWPAN Border Router)  having different constraints and requirements. Secure E2E connection only provisions secure communication channel and LLN (Low power and lossy networks) are still vulnerable to resource exhaustion, flooding, replay and amplification attacks. The authors have discussed two approaches to mitigate such attacks. First, mapping TLS or DTLS protocol to ensure end to end security at application layer which disallows 6LBR to gain access to data in transit. Second DTLS- DTLS tunnel is used to protect LLN.

Lightweight key pre-distribution schemes can be found in [33]. Such schemes are proposed for IoT. Improvement in resource efficiency of such algorithms is proposed in [30]. More efforts on optimization of the cryptographic operations in IoT security provisioning are proposed in [34] where the authors have shown the equivalence of the MMO problem to finding close vectors in a lattice. In [35] the authors proposed a new efficient ID based key establishment scheme. Identity based scheme consists of a node with an identifier and a Trusted Third Party (TTP) which provides the node in the network with secret keying material linked to the device identifier in a secure way. Secret keying material and the identity of the other node is used by other nodes to generate a common pairwise key for secure communication. Scheme put forward by the authors is efficient in terms of key computation time. In [36] the authors have proposed a key management service for BSN (body sensor networks). The key management service considers the resource inefficiency of IoT and low power devices.

The authors in [37] have presented security challenges of IoT communication. Architectural design for secure IP based IoT is reviewed by the authors. Lifecycle of an IoT device and its capabilities should be considered for designing a security architecture.  The architectural design should include the aspects of trusted third party and type of protocols applied. As another requirement, an architecture should scale from small-scale adhoc security domains to large-scale deployments. Lightweight protocols should be adopted within the architecture.  Another interested point raised by the authors is at which level (link, network or application layer) to base security in IoT. As each layer have different security requirements and communication patterns. If security is provision at application layer then the network is open to attacks. While focusing security on network or link layer introduce possible inter-application security threats.

Efforts in proposing lightweight security framework can be found in [38] where the proposed architecture provides a holistic security approach which contains lightweight authentication and authorization functionality on constrained smart objects. In [39] the authors proposed a lightweight framework comprising of DTLS, CoAP and 6LoWPAN protocols to provision end to end security for IoT. Efforts in evaluating the lightweight cryptographic framework can be found in [40], the authors proposed framework which assists embedded software engineer for selecting best cipher to match requirements of an application. Lightweight framework to provide access control in IoT can be found in [41]. The authors proposed a generic authorization framework for IoT devices. Evaluation of the proposed framework is also discussed.

Standard compliant security framework can be found in [42], the authors intend to make it applicable for future IoT paradigm. Another similar work that proposes an end-to-end security framework for IoT can be found in [43].

In [44] the authors proposed a security scheme for IoT based on established standards and prevailing Internet standards on a low-power hardware platform. Proposed security solution fails in preventing against routing attacks and is too heavy for low power devices. Another effort in evaluation of resource consumption during provisioning of security services can be found in [13]. The work lacks new proposal to deal with end-to-end security of IoT. The authors in [45] have evaluated secure communication mechanisms and compare them in terms of resource consumption. The work concludes with the best secure communication mechanism.

In [95], Hameed et al. proposed a security middleware for security weaknesses in NFC-based systems. The middleware in the initial stage is capable of detecting malicious NFC tags or smart posters with little effect on CPU and memory. The authors further extended their middleware with lightweight primitives to provide confidentiality and integrity support for arbitrary NFC applications [96] [97].

### 3.4 Open Issues and Future Directions

Most of the existing work have focused on optimizing the algorithmic steps in the cryptographic algorithms performing cryptographic operations. None of the current work consider framework of protocols that perform lightweight operations to secure the IoT network. Security provisioning over IoT is very challenging as compared to WSN due to heterogeneous nature of devices. Furthermore, they are deployed in unattended environments that are closer to humans than WSN nodes. In contrast to WSN IoT is expected to have IPv6, UDP and web support. Communication in IoT can be secured by (1) Lightweight security protocols proposed for constrained environments i.e. WSNs (2) Novel security protocols that meet the specific requirements of IoT (3) Established security protocols which are already existent in the Internet. The security protocols designed for WSN are not designed for IP network. In order to use them, IoT requires modification of WSN protocols and their provisioning in the Internet. Due to huge number of devices in the Internet, security solution require modification that is not practical in current Internet. Primary challenge that hinders applicability of Internet security solutions in IoT is that these solutions are not inherently designed for resource constrained devices but for standard computing machines which have sufficient energy resources, processing capability and storage space.

Apart from novel security solutions for IoT, emerging paradigm of SDN, with the potential of centralizing routing functionality enables central monitoring and reconfiguration of the network. This opens new possibilities of implementing lightweight cryptographic framework for IoT that runs light security protocols at SDN controller. Most of the major cryptographic functionalities are concentrated at the central controller. Hence, heavy cryptographic operations are offloaded to the central SDN controller that communicate with the IoT nodes.

## 4. Secure Routing and Forwarding in IoT

### 4.1 Motivation

IP based IoT inherits attack threats of IPv4. Some of these well known attacks are blackhole attacks, sybil, spoofing, smurfing, eavesdropping, neighbor discovery, man in the middle, rogue devices and fragmentation attacks. This means IoT is in need of same security measures as required for IPv4. As it is envisioned with IoT that the physical world will be connected with the Internet which leads to wide variety of security concerns. Attack threats not only include manipulation of information but actual control of devices in IoT network. With more electronic systems i.e. Modbus, SCADA becoming part of IP based systems, significant increase in attacks are expected. This adds new security threats as heterogeneous devices become part of IoT network.

In wireless mobile network a route is established when route information is transmitted from node to node until destination is found. Throughout this route maintenance phase, nodes are added or deleted. Furthermore, these nodes may unnecessarily delay transmission of control information, which usually is done by selfish or misbehaving nodes. During this phase of route setup and discovery several attacks are possible by malicious nodes in routing information. For example a certain node may introduce route table overflow attack by transmitting huge amount of false route information to neighboring nodes which cause neighbor's routing table to overflow. Due to such actions the table is filled with spurious routes and real routes are denied to occupy the routing table.

### 4.2 Challenges

Secure routing in IoT can be achieved by designing secure routing protocol. Such protocol should be able to securely establish a route and guarantee secure route among communicating nodes while isolating malicious nodes in the network. The protocol should self-stabilize which means it should be able to recover automatically from any kind of problem within a certain time without human involvement. Protocol should be able to isolate misbehaving nodes in the network so that disruption in routing process is done to minimal level. Lightweight computations should be performed by a secure routing protocol. Location privacy should be maintained for the IoT devices in the IoT network. Hence, for a secure routing protocol it should be able to maintain location privacy.

Current routing protocols for IoT are insecure as most IoT networks are self-organizing and usually operate without any human involvement. Hence, malicious nodes can be introduced in the IoT network.

### 4.3 Existing solution and Discussion

In this section, we discuss efforts in the direction of secure routing and forwarding in IoT.

IoT not only require provisioning of security services but often experience problems in routing and forwarding the data. Securing routing algorithm for IoT have become a crucial requirement. A comprehensive state of the art in securing routing for WSN can be found in [46]. The authors proposed a schematic taxonomy of key design issues in WSN routing protocols and define the design categorization factors for secure routing i.e. basic, essential and optional. Also a comparative study is performed on the basis of key design attributes, security objectives and attacks prevention which considered recent advancements in the area of secure WSN routing. Security aware routing protocols for adhoc networks can be found in [47]. The authors develop a generalize framework with open feedback and explicit representation of attributes and choices. With this users can adapt the security attributes in run time and talk terms for alternative routes which are calculated on the basis of cost benefit analysis of the performance penalties against offered protection in the scenario.

In [48] the authors have proposed an adaptive, flexible and lightweight scheme for protection of integrity and re-authentication based on hash chains which is often called as ALPHA. The proposed scheme enabled hop by hop and end to end integrity protection for multihop wireless networks. End to end integrity protection which is based on secret sharing will be replaced which cannot be authenticated by relays.

The authors in [49] proposed a secure and efficient Cost Assurance routing protocol for IoT (CASER). Routing in CASER is based on geography and do not rely on flooding to broadcast routing information in the network. It balances energy consumption and increase network lifetime. Furthermore, it sends messages by two routing strategies random walking and deterministic routing. Distribution of two strategies is decided by particular security requirements. The selection of two strategies is probabilistically controlled by assigning a probability to a variable representing security requirement that is dependent on the cost factor of the route. The authors presented quantitative security analysis of CASER. No software or component architectural details of CASER is discussed.

Advance security attacks in routing can be found in [50], the authors consider node capture attack where attacker capture a legitimate node and by extracting cryptographic keys it makes the captured node as malicious ones which run the malicious code. To attract the traffic the compromised node broadcast a fake RREQ with false hop count.

Secure multihop routing for IoT is proposed in [51], where multilayer parameters are embedded into the routing algorithm. It was shown by the authors that the proposed algorithm is suitable for IoT communication.

There are several efforts where researchers have proposed trust aware routing algorithms such as in [52] where authors claim to propose routing framework that have the attribute of lightweight and have high ability to resist various attacks. In [53] the authors proposed secure procedures for resource insufficient IoT devices. Comprehensive analysis of security capabilities of IoT can be found in [54], the analysis is performed by implementing and demonstrating famous routing attacks launched in 6LoWPAN network running RPL. Another work that gives a detail survey on security issues in IoT can be found in [55]. The authors discussed security measures that are adopted at heterogeneous layers of IoT architecture.

Other such efforts in proposing secure routing algorithm for IoT can be found in [56] [57] [58] [59][9]. In all of these work the authors have proposed trust aware secure routing algorithm for IoT. Detecting routing attacks in sensor networks using intrusion detection system can be found in [60]. The abnormal traffic behavior is detected using clustering algorithms that construct a model for normal traffic and conclude with abnormal traffic behavior.

In addition to efforts in secure routing there are proposals of detecting devastating attacks in IoT routing, such as in [61] an intrusion detection system to distinguish sinkhole attacks on the IoT routing services is proposed. Experimental evaluations show the effectiveness of the proposed idea.

### 4.4 Open Issues and Future Directions

Although there are various effort which have dealt with the problem of secure routing and forwarding. None of the work have considered performance of IoT network when secure routing mechanisms are incorporated. There are complex mechanisms such as the one that proposed IDS to detect attacks in the network. Such mechanism do not consider the resource limitation of IoT devices. However, lightweight IDS may help in detecting malicious activities in IoT network and mitigate routing attacks with IoT network. Novel designing of a lightweight IDS requires attention from the networking research community. Apart from lightweight IDS for IoT, regulating the IoT network from a central point can help in monitoring the state of the whole network. Paradigm such as SDN, which centralizes the control plane at the controller, can help in routing the data securely in IoT network. Hence, novel security solutions over SDN are required which route and forward IoT network data by preserving their integrity.

## 5. Robustness and Resilience Management in IoT

### 5.1 Motivation

IoT network constitute of heterogeneous devices where managing such kind of network is not an easy task. Lately, researchers have focused their attention towards Service Oriented Architecture (SOA) for the management of IoT [62], as it caters integration and management of diverse services. With the use of such paradigm a lightweight middleware can be constructed upon IoT devices providing abstraction of integrate-able and manageable IoT services. Developers and users of IoT devices are free of details on what and how devices are used and connected. Faults in IoT applications is not tolerable, as system failures may disrupt user's every day activities or even lead to danger in lives. To worsen the situation SOA is prone to all the faults related to distributed systems [63]. Hence, SOA based middleware for IoT is subjected to all the inherent problems of distributed systems. Besides normal faults in IoT devices, such faults may occur due to DoS attacks on IoT devices and services disrupting IoT services to the application users.

## 5.2 Challenges

Timely recovery from failures becomes critical in IoT network. Prolong disruption in IoT services may lead to life threatening situation especially for disaster management applications. Hence, it is required that the resource management middleware designed for IoT network should timely detect failures and resolve the situation. There are various possible solutions to resolve IoT device failures. One of the possible solution is to replicate resources [64] and deploy them in the same environment. This solution are may be costly, as it require duplication of resources.

## 5.3 Existing solution and Discussion

In this section, we discuss efforts in the direction of ensuring robustness in IoT network.

In [65] the authors addressed the problem of misbehaving node in the network. They proposed a preliminary description of protocol ECoSec (Efficient Cooperative Security) which controls the admission and revocation of nodes by collaborating with other nodes by two voting procedures. Trust management is also looked upon by the researchers to provide resilience in IoT. In [66] the authors have analyzed the proposed protocol ECoSec (Efficient Cooperative Security Protocol) operation and parameters for node agreement that is performed for admission voting and revocation information. Other work done in handling such issues by context awareness and intelligence can be found in [67] [68]. In [69] the authors discussed resilience management in IoT. AI based approaches to provide fault tolerance in IoT can be found in [70] where the authors proposed hybrid cross layer and fault tolerant routing protocol based on learning automata. The algorithm dynamically adapt to the dynamic environment and then chooses optimal action. Also, the algorithm adopts fault tolerant routing which is energy aware. Energy is conserved by sleep algorithm that is coordinated by a dynamic and adaptive scheduling algorithm.

Efforts in fault management can be found in [71] where authors put forward a fault management structure that is layered for diverse IoT networks. In order to realize efficient end-to-end transmission, fault detection and location fuzzy cognitive maps theory is introduced. The authors do not evaluate overhead of the layered architecture. In [72] distributed fault tolerance is developed and implemented which will configure itself based on user policy and requirements. Overhead of incorporating middleware and framework for IoT and M2M is not discussed in this work. Further, the effectiveness of distributed mechanism as compared to centralized mechanism is also not discussed by the authors.

In another work [73] a self-learning based sensor fault detection for monitoring industrial IoT is put forward by the authors. Responsiveness of the self-learning module is not evaluated and discussed. In [74] the authors proposed a network management framework for WSN, which is self-optimizing, and fault tolerant. Cost associated with message passing middleware is not discussed in this work. Researchers have considered cloud-computing framework to provide fault tolerance to WSN such as in [75]. Evaluation of configuring failures in sensor network is not discussed in this work as well. Similar effort can be found in [76] where the authors present a cloud-based framework to evaluate failures in sensor network.

In [77], the authors propose network management protocol for WSN to manage failures in the network. Similarly, in [78] the authors proposed novel architecture for scalability and fault tolerance in healthcare. Fault tolerance is attained by backup routing between nodes. Other such work related to managing WSN can be found in [79]. More work on management of M2M can be found in [80] where minimum requirements for M2M network management is presented along with standardization activities.

## 5.4 Open Issues and Future Directions

People have approached the problem of ensuring robustness in IoT network by proposing protocols and network management framework. Faults in IoT network can occur due to either network attacks or depletion of energy. Efforts in tackling faults are numerous and most of them have not considered the resource constraint nature of IoT devices. Centralizing the network view can ensure failures over IoT network to be controlled and provision fault tolerant routing. As the decisions of routing will be concentrated on the controller, it will be possible to detect faults centrally. By detecting faults, decisions to divert the traffic to an alternative server or path will be carried out at the controller. Creative solutions that detect faults in a timely manner are required so that actions can be taken promptly to handle the situation by suggesting alternate possibilities.

# 6. Denial of Service and Insider Attack Detection in IoT

## 6.1 Motivation

Denial of Service (DoS) attacks have devastating effects in IoT applications [81]. In IoT applications, availability of IoT service and devices is an important factor. DoS attacks make the IoT services unavailable, thus disrupting their normal operations. DDoS attacks are normally launched in a coordinated manner from multiple attackers at the same time and their detection before the services become unavailable is quite difficult.

With IoT becoming an integral part of business applications. Businesses face remarkable challenge of understanding and addressing risks of protecting themselves from range of insider attacks. These attacks are usually launched by the use of devices that are unknown; remain undetectable and unmanaged by the IoT applications.

## 6.2 Challenges

As DoS attacks are difficult to detect before the attack is launched. Efficient DoS detection solutions are required. There exist various proposals where DDoS is dealt for traditional Internet such as in [82]. Detection of DDoS in IoT is a challenging issue as IoT network and traffic characteristics are quite different from traditional network. Due to limitation of IoT devices, resource efficient DDoS detection and countermeasure techniques are required. Such techniques can be centralized such that based on monitoring the traffic in the IoT network centrally. Certain probabilistic techniques can help in inferring possibility of DDoS attacks. On the other hand, such techniques can be distributed where multiple IoT devices collaboratively infer possibility of DDoS attack in IoT network.

In order to prevent insider attacks in IoT it is required to authorize IoT nodes becoming part of the IoT network. Techniques for detecting insiders in IoT network should be efficient and react in a timely manner. Otherwise, devastating situation may arise as these insiders may leak confidential data by compromising nodes within the network or disrupt operation of the IoT network by launching attacks such as DDoS attacks.

## 6.3 Existing Solution and Discussion

In this section we discuss efforts in countering DDoS and insider attacks in IoT.

Insider attack have received attention from researchers such as in [83] the authors proposed a mechanism that manages the network using a node, which monitor the network constantly. The proposed algorithm works by maintaining a dynamic threshold. The Threshold is adjusted by the view of overall packet loss situation in real time. This results in decrement of the detection rate due to loss and rate associated with false alarm. The authors put trust mechanism based secure routing protocol in [84]. They proposed an algorithm that investigate neighborhood activities based on the mechanism of spatial correlation and requires no knowledge of malicious sensor. It is important as the pre-hand knowledge of sensor causes excess training overhead and discusses a grave strain in which attack behaviors alter dynamically. In [85] the authors proposed a rule based anomaly detection system called RADS. The proposed idea revolves around detecting sybil attacks in 802.15.4 like WSNs by monitoring.

Efforts addressing DDoS attacks in IoT are discussed as follows. In [86] the authors put forward a framework that can be spread out in existing network and can prevent forged messages that are broadcasted in the entire network. The filters are used to actually verify fake messages. Some of the nodes in the network have high processing and battery power than the other nodes. The nodes are called as adjunct nodes that are used to monitor the state of network and perform appropriate actions when required. The authors in [87] aims to save WSN from DDoS attack using a mechanism which utilizes profile for provisioning security against various attack. Sensor nodes monitor the surrounding environment and deliver acquired data to the sink node for profiling. Profile based protection scheme (PPS) is used to supervise the activities performed in the network. A comprehensive taxonomy of DoS attacks in WSN can be found in [88]. The taxonomy identifies attacker, attack victim and vulnerabilities.

A detail description of IDS in application of IoT i.e. SCADA can be found in [89] and it has briefly discussed the history of research in IDS techniques. 6LoWPAN based IoT when subjected to denial of service attack often experience devastating situation. In [90] the author's proposed DoS detection architecture for 6LoWPAN based IoT. The authors do not evaluate cost of communication among components of the proposed architecture and overhead. Moreover, being a centralized architecture it is subjected to single point of failure. Variants of broadcast protocol i.e. TESLA for IoTs which is DoS tolerant can be found in [91].

In [92] the authors proposed an artificial intelligent based approach to counter attack DDoS by proposing learning automata based preventive scheme. SoA based framework is used for IoT due to its huge potential for huge number of applications. A cross layer model for DDoS mitigation is used. IoT's are typically resource constraint hence communication among layers incurs cost. The authors do not evaluate effectiveness of cross layer model. Learning automata mechanism is heavy to implement which might not be feasible for huge network with various types of IoT devices.

In [93] the authors proposed an IDS framework for IoT. Monitoring system and detection engine are the components of the proposed framework. Evaluation of the proposed architecture with the increasing size of 6LoWPAN is not discussed. Communication cost of IDS framework will increase with the increasing size of the 6LoWPAN. Latest work in IDS for IoT can be found in [94] where possible attacks in IoT is discussed and along with lacking in current IDS for IoT. Different categories of IDS for IoT are discussed and various existing types of IDS for conventional Internet are highlighted which are to be evaluated for RPL network.

**6.4 Open Issues and Future Directions**

Most of the proposed framework for tackling DDoS and insider attack is based on monitoring system and detection engine. Implementing detection engine over IoT network is resource consuming as they are based on AI algorithms. Hence, novel lightweight solutions for detecting DoS attacks is required. Apart from novel lightweight solutions emerging paradigm of SDN enables monitoring of network state from a central point called controller. By monitoring flows at the controller, it is possible to implement algorithms to detect DDoS attacks and malicious activities such as insider attack [29]. This will also offload the tasks of defeating DDoS attacks from IoT devices to resource sufficient device that hosts SDN controller possibly the gateway device connecting IoT devices. A good hybrid solution would be to integrate IoT gateways with emerging SDN based solutions that are capable of efficiently detecting [98] and mitigating [99] DDoS in conventional IP networks.

## 7. Conclusion

In this paper, we have categorized and discussed state of the art work done in ensuring security in IoT network. Efforts in privacy provisioning, light weight cryptographic framework, secure routing and forwarding, robustness and resilience management, denial of service and insider attack detection are discussed comprehensively. Privacy is crucial in IoT especially as the characteristics of such network is different than the typical Internet network. Such issues and requirements are identified and discussed in this paper. Besides privacy for ensuring security in IoT network, lightweight cryptographic primitives are required which are suited for IoT network. All the efforts in this direction are compiled and future actions are discussed.

In order to preserve privacy, context aware techniques, lightweight protocols are proposed and most lately virtualization techniques are used to maintain integrity of the data. For lightweight cryptographic primitives novel solutions are required which should consume limited resources of an IoT mote. Apart from that SDN solution offers to implement lightweight cryptographic solutions over IoT with the assistance of centralized routing carried at the SDN controller. IoT network experiences failures due to IoT nodes being subjected to heterogeneous kind of network attacks. Efforts in this direction are discussed with future insight. Faulty nodes within IoT network can be experienced due to Denial of Service attacks launched by multiple coordinated nodes. Furthermore, such faults are prevalent due to frequent insider attack within the IoT network. To realize fault tolerance in IoT centralize monitoring of the network state is required in order to timely react to counter faulty nodes with in the network. Virtualization technology like SDN offers centralize monitoring of the network which can assist in suggesting alternative servers or path to ensure consistent provisioning of service. As far as DDoS in IoT is concerned lightweight detection engine suitable for IoT is required to detect and mitigate DDoS in a timely manner. Centralize monitoring enabled by SDN can assist in detecting DDoS and mitigate them with in an IoT network.

For all of the security requirements there is a need of a centralized management framework which can provision all the discussed security issues and requirements with in IoT network. SDN is a hot candidate which provides central configuration of the network by the controller which manages the network. Initial efforts in this direction can be found in [29]. There are still lot of opportunities and issues which need to be dealt with in order to realize a comprehensive centralized management framework for provisioning security over IoT. SDN need to be studied thoroughly so that it can be customized to provision management services over IoT network.